# Semiquantum private comparison based on Bell states without quantum measurements from the classical user

**Mao-Jie Geng[1], Xia Li[1] and Tian-Yu Ye[1]***

[1] College of Information & Electronic Engineering, Zhejiang Gongshang University, Hangzhou 310018, P.R.China

E-mail: yetianyu@zjgsu.edu.cn

### Abstract

In this paper, we propose a novel semiquantum private comparison (SQPC) protocol based on Bell states, which enables one quantum user and one classical user to compare the equality of their private inputs with the help of a semi-honest quantum third party (TP). TP is assumed to be semi-honest in the sense that she may take all possible attacks to steal users’ private inputs except conspiring with anyone. The security analysis validates that our protocol can resist not only the attacks from internal participants but also the attacks from an external eavesdropper. Besides, our protocol only asks TP to perform Bell basis measurements but doesn’t need quantum entanglement swapping; and it releases the classical user from conducting quantum measurements and having a quantum memory. Moreover, our protocol can take advantage over previous SQPC protocols based on Bell states in qubit efficiency. Finally, our protocol can be generalized into its counterpart of the collective-dephasing noise quantum channel.

### 1. Introduction

In the year of 1984, Bennett and Brassard [1] proposed the first quantum cryptography protocol, namely the famous BB84 quantum key distribution (QKD) protocol. Hereafter, quantum cryptography has been developed rapidly. In the year of 2009, Yang and Wen [2] proposed the novel concept named as quantum private comparison (QPC), in order to accomplish the goal that the equality of private inputs from two different users is compared through the quantum means under the condition that none of these private inputs are leaked out. Subsequently, scholars put forward a series of QPC protocols by using different quantum resources, such as single particles [3,4], Bell states [5-8], GHZ states [9,10], $\chi$ -type entangled states [11], five-qubit entangled states [12], six-qubit entangled states [13], *etc*.

In the years of 2007 and 2009, Boyer *et al*. [14,15] respectively proposed two pioneering semiquantum key distribution (SQKD) protocols by using polarized single photons to claim the birth of semiquantum cryptography. Later, Zou *et al.* [16] proposed a novel SQKD protocol to release the classical user from quantum measurements; Ye *et al.* [17,18] put forward two novel SQKD protocols with single photons in both polarization and spatial-mode degrees of freedom. According to the two works of Refs.[14,15], in the realm of semiquantum cryptography, it is generally believed that the classical user is limited to the following operations: (a) transmitting the qubits without disturbance; (b) measuring the qubits with the $Z$ basis (i.e., $\{|0\rangle, |1\rangle\}$); (c) producing the qubits in the $Z$ basis; and (d) scrambling the qubits with delay lines. In the year of 2016, Chou *et al.* [19] proposed the first semiquantum private comparison (SQPC) scheme based on Bell states and quantum entanglement swapping. Compared with QPC, SQPC releases the classical communicant from the preparation and measurement of quantum superposition states and quantum entangled states, and is advantageous for the classical communicant to reduce the burdens of quantum state preparation and measurement. Since then, scholars have proposed numerous SQPC protocols with different quantum states. For example, Refs.[20-24] utilized Bell states to propose different SQPC protocols; Ref.[25] put forward a measure-resend SQPC protocol by using two-particle product states; Refs.[26,27] proposed different SQPC protocols by using single photons; Ref.[28] put forward a SQPC scheme based on Greenberger-Horne-Zeilinger (GHZ) class states; Ref.[29] suggested a novel SQPC scheme based on entanglement swapping of four-particle cluster states and Bell states; Ref.[30] put forward a multi-party SQPC scheme based on the maximally entangled GHZ-type states; Ref.[31] designed a novel SQPC scheme with W states; Ref.[32] put forward a novel SQPC scheme via cavity QED by using two-atom product states; Ref.[33] constructed a feasible SQPC scheme based on entanglement swapping of Bell states. Each of Refs.[19-33] aims to compare the equality of two classical users' private inputs. Different from Refs.[19-33], the SQPC schemes of Refs.[34-42] can compare the size relationship of classical users' private inputs. Ref.[34], Ref.[38] and Ref.[40] adopted $d$ -dimensional Bell states to design the SQPC protocol which can compare the size relationship of two classical users' private inputs; Ref.[35], Ref.[36], Ref.[39] and Ref.[41] constructed the SQPC protocols of size relationship by using $d$ -dimensional single-particle states; Ref.[37] designed the SQPC protocols of size relationship by using $d$ -dimensional GHZ states; Ref.[42] designed a novel SQPC protocol with $d$ -dimensional $N+1$ -particle maximal entangle states. However, each of the above SQPC protocols in Refs.[19-42] compares classical users' private inputs. As far as we know, at present, there is no SQPC protocol which can compare one quantum user's private input and one classical user's private input.

Based on the above analysis, in this paper, we propose a novel SQPC protocol based on Bell states first, which can correctly compare the equality of one quantum user's private input and one classical user' private input with the help of a semi-honest quantum third party (TP). Besides, our protocol doesn't need quantum entanglement swapping, doesn't require the classical user to conduct quantum measurements and have a quantum memory, and only needs TP to perform Bell basis measurements. Moreover, the qubit efficiency of our protocol exceeds that of each of previous SQPC protocols based on Bell states [19-24]. Finally, we generalize our protocol into its counterpart of the collective-dephasing noise quantum channel.

## 2.Protocol description

Suppose that Alice is the user equipped with unlimited quantum capabilities, while Bob is the user only having limited quantum capabilities. Alice and Bob hold private inputs $X=(x_1,x_2,\dots,x_n)$ and $Y=(y_1,y_2,\dots,y_n)$, respectively, where $x_i,y_i\in\{0,1\}$, $i=1,2,\dots,n$. They want to compare the equality of their private inputs with the help of a semi-honest TP, who is assumed to have unlimited quantum capabilities. Relying on Ref.[43], "semi-honest" means that TP may take all possible attacks to obtain two users' private inputs except conspiring with anyone. The specific steps of the proposed SQPC protocol are as follows, which are further shown in Fig.1 for clarity.

Step 1: Alice (Bob) generates two $n$ -bit random number sequences $R_{A1}=(r_{A1}^1,r_{A1}^2,\dots,r_{A1}^n)$ ($R_{B1}=(r_{B1}^1,r_{B1}^2,\dots,r_{B1}^n)$) and $R_{A2}=(r_{A2}^1,r_{A2}^2,\dots,r_{A2}^n)$ ($R_{B2}=(r_{B2}^1,r_{B2}^2,\dots,r_{B2}^n)$) by applying a random number generator, where $r_{A1}^i,r_{A2}^i\in\{0,1\}$ ($r_{B1}^i,r_{B2}^i\in\{0,1\}$) and $i=1,2,\dots,n$. Thereafter, Alice (Bob) encrypts $X$ ($Y$) with $R_{A1}$ ($R_{B1}$) and obtains $X'=\{x_1',x_2',\dots,x_n'\}=\{x_1\oplus r_{A1}^1,x_2\oplus r_{A1}^2,\dots,x_n\oplus r_{A1}^n\}$ ( $Y'=$

$\{y'_1, y'_2, \dots, y'_n\} = \{y_1 \oplus r^1_{B1}, y_2 \oplus r^2_{B1}, \dots, y_n \oplus r^n_{B1}\}$ ), where $\oplus$ is the bitwise XOR operation. Besides, Alice agrees on with Bob in advance to place $R_{A2}$ and $R_{B2}$ in the same positions of $X'$ and $Y'$, so as to form new sequences $A = \{a_1, a_2, \dots, a_{2n}\}$ and $B = \{b_1, b_2, \dots, b_{2n}\}$, respectively. For example, let $A = \{x'_1, x'_2, \dots, x'_n, r^1_{A2}, r^2_{A2}, \dots, r^n_{A2}\}$ and $B = \{y'_1, y'_2, \dots, y'_n, r^1_{B2}, r^2_{B2}, \dots, r^n_{B2}\}$. Moreover, Alice and Bob share two $n$-bit private keys, $K = \{k^1, k^2, \dots, k^n\}$ and $K_{AB} = \{k^1_{AB}, k^2_{AB}, \dots, k^n_{AB}\}$, via the SQKD protocol in Ref.[16] beforehand, where $k^i, k^i_{AB} \in \{0,1\}$ and $i = 1,2, \dots, n$. According to $K_{AB}$, Alice and Bob produce another $4n$-bit private key $K'_{AB} = \{k'^1_{AB}, k'^2_{AB}, \dots, k'^{4n}_{AB}\} = \{k^1_{AB}, k^2_{AB}, \dots, k^n_{AB}, k^1_{AB}, k^2_{AB}, \dots, k^n_{AB}, k^1_{AB}, k^2_{AB}, \dots, k^n_{AB}, k^1_{AB}, k^2_{AB}, \dots, k^n_{AB}\}$.

Step 2: TP prepares $4n$ Bell states, each of which is randomly selected from one of the four states $\{|\phi^+\rangle, |\phi^-\rangle, |\psi^+\rangle, |\psi^-\rangle\}$, where $|\phi^\pm\rangle = \frac{1}{\sqrt{2}}(|00\rangle \pm |11\rangle)$ and $|\psi^\pm\rangle = \frac{1}{\sqrt{2}}(|01\rangle \pm |10\rangle)$. Then, she splits all first particles and all second particles into sequences $S_A$ and $S_B$, respectively. Finally, TP sends the particles of $S_A$ ($S_B$) to Alice (Bob) one by one. Take notes that after TP sends the first particle to Alice (Bob), she sends a particle only after receiving the prior one.

Step 3: After receiving $s^j_A$ ($s^j_B$) from TP, Alice (Bob) performs the corresponding operation on it according to $k'^j_{AB}$, where $s^j_A$ ($s^j_B$) is the $j$th particle of $S_A$ ($S_B$), $j = 1,2, \dots, 4n$. That is, when $k'^j_{AB} = 0$, Alice (Bob) performs the SIFT operation on it; and when $k'^j_{AB} = 1$, Alice (Bob) performs the CTRL operation on it. Here, the CTRL operation means to directly return the received particle to TP, while the SIFT operation means to perform the corresponding unitary operation on the received particle and then send the resulted particle to TP. The rule of performing the unitary operation on the $l$th SIFT particle is that: when $a_l = 0$ ($b_l = 0$), Alice (Bob) performs $I = |0\rangle\langle 0| + |1\rangle\langle 1|$ on the $l$th SIFT particle; and when $a_l = 1$ ($b_l = 1$), Alice (Bob) performs $\sigma = |0\rangle\langle 1| + |1\rangle\langle 0|$ on the $l$th SIFT particle. Here, $l = 1,2, \dots, 2n$. For convenience, $S_A$ ($S_B$) after Alice's (Bob's) operations is represented by $S'_A$ ($S'_B$).

Step 4: After TP obtains the particles returned from Alice and Bob, she measures the particles of the same positions in $S'_A$ and $S'_B$ with Bell basis and writes down the corresponding measurement results. Alice and Bob tell TP the positions where they chose the CTRL operations. TP derives a bit sequence $C = \{c_1, c_2, \dots, c_{2n}\}$ from her Bell basis measurement results on the SIFT particles of the same positions in $S'_A$ and $S'_B$: when her measurement result is same as the initial prepared state, TP sets $c_l = 0$; otherwise, she sets $c_l = 1$. Here, $l = 1,2, \dots, 2n$. For clarity, the relationships among different parameters corresponding to SIFT particles of the same positions in $S'_A$ and $S'_B$ are summarized in Table 1.

Table 1 Relationships among different parameters corresponding to SIFT particles of the same positions in $S_A'$ and $S_B'$

| Initial prepared states | $a_l$ | $b_l$ | Alice's unitary operation | Bob's unitary operation | TP's measurement result | $c_l$ |
|---|---|---|---|---|---|---|
| $\|\phi^+\rangle$ | 0 | 0 | $I$ | $I$ | $\|\phi^+\rangle$ | 0 |
| $\|\phi^+\rangle$ | 0 | 1 | $I$ | $\sigma$ | $\|\psi^+\rangle$ | 1 |
| $\|\phi^+\rangle$ | 1 | 0 | $\sigma$ | $I$ | $\|\psi^+\rangle$ | 1 |
| $\|\phi^+\rangle$ | 1 | 1 | $\sigma$ | $\sigma$ | $\|\phi^+\rangle$ | 0 |
| $\|\phi^-\rangle$ | 0 | 0 | $I$ | $I$ | $\|\phi^-\rangle$ | 0 |
| $\|\phi^-\rangle$ | 0 | 1 | $I$ | $\sigma$ | $\|\psi^-\rangle$ | 1 |
| $\|\phi^-\rangle$ | 1 | 0 | $\sigma$ | $I$ | $\|\psi^-\rangle$ | 1 |
| $\|\phi^-\rangle$ | 1 | 1 | $\sigma$ | $\sigma$ | $\|\phi^-\rangle$ | 0 |
| $\|\psi^+\rangle$ | 0 | 0 | $I$ | $I$ | $\|\psi^+\rangle$ | 0 |
| $\|\psi^+\rangle$ | 0 | 1 | $I$ | $\sigma$ | $\|\phi^+\rangle$ | 1 |
| $\|\psi^+\rangle$ | 1 | 0 | $\sigma$ | $I$ | $\|\phi^+\rangle$ | 1 |
| $\|\psi^+\rangle$ | 1 | 1 | $\sigma$ | $\sigma$ | $\|\psi^+\rangle$ | 0 |
| $\|\psi^-\rangle$ | 0 | 0 | $I$ | $I$ | $\|\psi^-\rangle$ | 0 |
| $\|\psi^-\rangle$ | 0 | 1 | $I$ | $\sigma$ | $\|\phi^-\rangle$ | 1 |
| $\|\psi^-\rangle$ | 1 | 0 | $\sigma$ | $I$ | $\|\phi^-\rangle$ | 1 |
| $\|\psi^-\rangle$ | 1 | 1 | $\sigma$ | $\sigma$ | $\|\psi^-\rangle$ | 0 |

In order to check the transmission security of CTRL particles, for the positions where Alice and Bob chose the CTRL operations, TP checks whether her measurement results on the CTRL particles of the same positions in $S_A'$ and $S_B'$ are same as the initial states prepared by herself. If all results are positive, the protocol will be continued; otherwise, the protocol will be halted.

In order to check the transmission security of SIFT particles, Alice and Bob publish the positions where they performed the unitary operations according to $R_{A2}$ and $R_{B2}$, respectively. Then, Alice and Bob require TP to publish the initial prepared states of these chosen positions and the corresponding measurement results, while TP requires Alice and Bob to publish the values of $R_{A2}$ and $R_{B2}$, respectively. Then, Alice, Bob and TP check whether the initial prepared states of these chosen positions, TP's corresponding measurement results and Alice and Bob's corresponding unitary operations are correctly related or not. If all results are positive, the protocol will be continued; otherwise, the protocol will be halted.

Step 5: TP drops out the bits in $C$ corresponding to $R_{A2}$ and $R_{B2}$, and obtains the new bit sequence $C' = \{c_1', c_2', \ldots, c_n'\}$. Note that the bits in $C'$ are corresponding to $X'$ and $Y'$.

Step 6: Alice (Bob) informs TP of the value of $r_{A1}^i \oplus k^i$ ( $r_{B1}^i \oplus k^i$ ), where $i = 1,2,\ldots,n$ . Afterward, TP calculates $m_i = \left(r_{A1}^i \oplus k^i\right) \oplus \left(r_{B1}^i \oplus k^i\right) \oplus c_i'$ for $i = 1,2,\ldots,n$. TP concludes that $X \neq Y$ as long as she finds out $m_i \neq 0$ for certain $i$; otherwise, TP concludes that $X = Y$. Finally, TP tells Alice and Bob the final comparison result of $X$ and $Y$.

Until now, it has finished the description of the proposed SQPC protocol. According to Ref.[44], the unitary operation $\sigma$ can be regarded to be classical. In the proposed protocol, classical Bob needs to receive qubits, apply the unitary operation $\sigma$, send out qubits, prepare fresh particles in the $Z$ basis (according to Ref.[16]), reorder particles via different delay lines (according to Ref.[16]), but is not required for quantum measurements or a quantum memory; quantum Alice needs to prepare fresh particles in the $X$ basis and $Z$ basis (according to Ref.[16]), receive qubits, apply the unitary operation $\sigma$, and send out qubits, but is not required for quantum measurements or a quantum memory; quantum TP needs to prepare Bell states, implement Bell basis measurements, send out qubits, receives qubits, and keep qubits in a quantum memory.

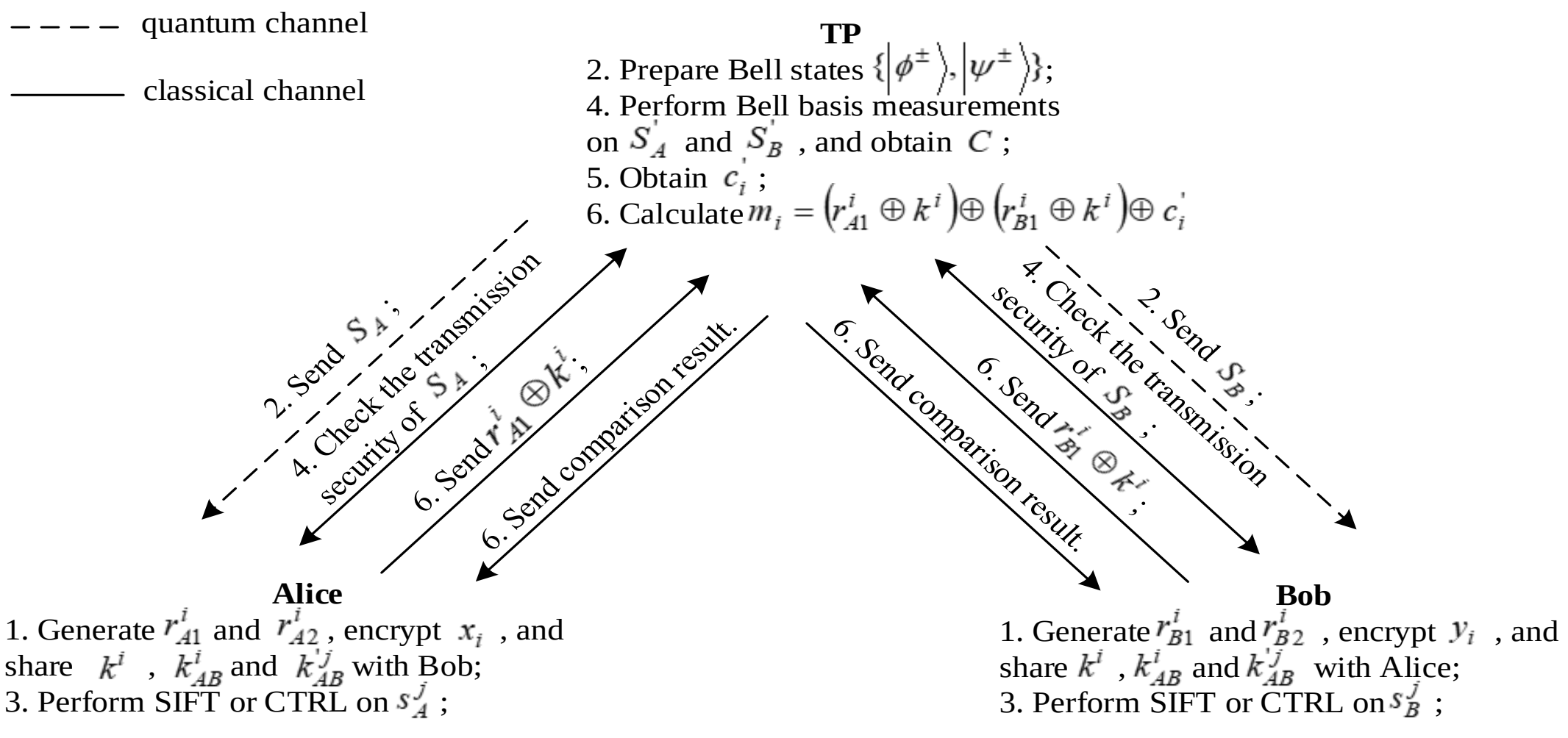

Fig.1 The flowchart of the proposed SQPC protocol

## 3. Correctness analysis

In Step 1, Alice (Bob) uses a random number generator to generate $R_{A1}$ ( $R_{B1}$ ) and $R_{A2}$ ( $R_{B2}$ ). Then, Alice (Bob) obtains $X'$ ($Y'$) by encrypting $X$ ($Y$) with $R_{A1}$ ($R_{B1}$). Afterward, Alice agrees on with Bob in advance to place $R_{A2}$ and $R_{B2}$ in the same positions of $X'$ and $Y'$ to form $A$ and $B$, respectively. Then, Alice and Bob share $K_{AB}$ via the SQKD protocol in Ref.[16] beforehand. According to $K_{AB}$, Alice and Bob produce $K_{AB}'$.

In Step 2, TP prepares $4n$ Bell states randomly in $\{|\phi^+\rangle, |\phi^-\rangle, |\psi^+\rangle, |\psi^-\rangle\}$. Then, she obtains $S_A$ and $S_B$, and sends the particles of $S_A$ ($S_B$) to Alice (Bob) one by one.

In Step 3, after receiving $s_A^j$ ($s_B^j$) from TP, Alice (Bob) performs the SIFT operation or the CTRL operation on it according to $k_{AB}'^j$, where $s_A^j$ ($s_B^j$) is the $j$th particle of $S_A$ ($S_B$), $j = 1,2,\dots,4n$. The rule of performing the unitary operation on the $l$th SIFT particle is that: when $a_l = 0$ ($b_l = 0$), Alice (Bob) performs $I = |0\rangle\langle 0| + |1\rangle\langle 1|$ on the $l$th SIFT particle; and when $a_l = 1$ ($b_l = 1$), Alice (Bob) performs $\sigma = |0\rangle\langle 1| + |1\rangle\langle 0|$ on the $l$ th SIFT particle. Here, $l = 1,2,\dots,2n$. $S_A$ ($S_B$) after Alice's (Bob's) operations is represented by $S_A'$ ($S_B'$).

In Step 4, TP produces $C$ from her Bell basis measurement results on the SIFT particles of the same positions in $S_A'$ and $S_B'$. According to Table 1, we can clearly obtain that $c_i' = x_i' \oplus y_i'$, where $i = 1,2,\dots,n$. Besides, we can know $x_i' = x_i \oplus r_{A1}^i$ and $y_i' = y_i \oplus r_{B1}^i$ from Step 1. Hence, we have

$$\begin{aligned} m_i &= (r_{A1}^i \oplus k^i) \oplus (r_{B1}^i \oplus k^i) \oplus c_i' \\ &= r_{A1}^i \oplus r_{B1}^i \oplus c_i' \\ &= r_{A1}^i \oplus r_{B1}^i \oplus (x_i' \oplus y_i') \\ &= r_{A1}^i \oplus r_{B1}^i \oplus (x_i \oplus r_{A1}^i) \oplus (y_i \oplus r_{B1}^i) \\ &= x_i \oplus y_i. \end{aligned} \tag{1}$$

According to Eq.(1), $m_i = 0$ means that $x_i = y_i$. It can be drawn the conclusion now that the output of the proposed protocol is correct.

## 4.Security analysis

### *4.1 Outside attacks*

An external eavesdropper, Eve, wants to obtain something useful about users' private inputs by launching various famous attacks, such as the Trojan horse attacks, the intercept-resend attack, the measure-resend attack. However, Eve can be inevitably detected.

(1) The Trojan horse attacks

Because the particles of $S_A$ ($S_B$) are transmitted from TP to Alice (Bob) and back to TP, we need to think over the invisible photon eavesdropping attack [45] and the delay-photon Trojan horse attack [46,47] from Eve. It has been verified that a wavelength filter and a photon number splitter can be added in front of Alice's (Bob's) device to resist the invisible photon eavesdropping attack and the delay-photon Trojan horse attack, respectively [47,48].

(2) The intercept-resend attack

In Step 1, TP sends the particles of $S_A$ ($S_B$) to Alice (Bob). Eve intercepts the particles sent from TP to Alice (Bob), and sends Alice (Bob) the false ones she prepared in advance in the $Z$ basis; after Alice's (Bob's) operations, Eve measures the particles sent out from Alice (Bob) with the $Z$ basis, in hoping of getting something useful about Alice's (Bob's) private input, and sends the resulted states to TP. When Alice and Bob tell TP the positions where they chose the CTRL operations, Eve may hear of them. In this way, Eve may know the positions where Alice and Bob chose the SIFT operations. Through this intercept-resend attack, Eve can decode out $A$ ($B$) from the initial prepared states of the fake particles corresponding to Alice's (Bob's) SIFT operations and her measurement results on the corresponding particles sent out from Alice (Bob) before being detected. However, she still cannot obtain $X$ ($Y$) before being detected, because Eve has no knowledge about $R_{A1}$ ($R_{B1}$) and the position of $X'$ ($Y'$) in $A$ ($B$).

In the following, we validate that this kind of attack from Eve can be discovered by Alice, Bob and TP inevitably. For example, assume that an initial pair of particles prepared by TP is $|\varphi^+\rangle$ and that the two fake particles prepared by Eve are both in the state of $|0\rangle$. After Alice and Bob perform their operations, TP executes Bell basis measurement on this pair of returned particles. When both Alice and Bob choose the CTRL operations, TP's measurement result is randomly in the state of $|\varphi^+\rangle$ or $|\varphi^-\rangle$; as a result, Eve is detected with the probability of $\frac{1}{2}$, as TP's measurement result should be $|\varphi^+\rangle$ when no attack happens. When both Alice and Bob choose the SIFT operations, Eve is also detected with the probability of $\frac{1}{2}$ if this pair of particles is chosen for security check.

(3) The measure-resend attack

In the process of TP transmitting the particles of $S_A$ ($S_B$) to Alice (Bob), Eve intercepts them, measures them with the $Z$ basis and sends the resulted states to Alice (Bob); after Alice's (Bob's) operations, Eve transmits the resulted states to TP after measuring the particles sent out from Alice (Bob) with the $Z$ basis. Eve may know the positions where Alice and Bob chose the SIFT operations after they tell TP the positions where they chose the CTRL operations. Through this measure-resend attack, Eve can deduce $A$ ($B$) from her measurement results on the SIFT particles from TP to Alice (Bob) and back to TP before being detected. Unfortunately, Eve still cannot know $X$ ($Y$) before being detected, due to lack of $R_{A1}$ ($R_{B1}$) and the position of $X'$ ($Y'$) in $A$ ($B$).

In addition, Alice, Bob and TP can successfully detect Eve's this kind of attack. For example, assume that an initial pair of particles prepared by TP is $|\varphi^+\rangle$. After Eve intercepts these two particles and measures them with the $Z$ basis, they are collapsed randomly into $|0\rangle|0\rangle$ or $|1\rangle|1\rangle$. Without loss of generality, assume that they are collapsed into $|0\rangle|0\rangle$. After Alice and Bob perform their operations, TP performs Bell basis measurement on this pair of returned particles. When both Alice and Bob choose the CTRL operations, TP obtains $|\varphi^+\rangle$ or $|\varphi^-\rangle$ with the same probability; hence, Alice, Bob and TP can detect Eve with the probability of $\frac{1}{2}$, as TP's measurement result should be $|\varphi^+\rangle$ when no attack happens.

When both Alice and Bob choose the SIFT operations, Alice, Bob and TP can also detect Eve with the probability of $\frac{1}{2}$ if this pair of particles is chosen for security check.

### *4.2 Participant attacks*

In 2007, Gao *et al.* [49] reminded that participant attacks must be given more concerns to, due to their strong powers. Therefore, in the proposed protocol, we need to pay special attention to the attacks launched by Alice, Bob or TP.

(1) The participant attack from Alice or Bob

In the proposed protocol, Alice possesses unlimited quantum capabilities, while Bob only has limited quantum capabilities. Hence, it can be thought that Alice is more powerful than Bob. In this regard, here analyzes the participant attack from Alice first.

Firstly, we analyze the intercept-resend attack from Alice. Alice introduces no disturbances on the particles of $S_B$ sent out from TP corresponding to the CTRL operations, but intercepts the particles of $S_B$ sent out from TP corresponding to the SIFT operations and uses the fake particles prepared by her in the $Z$ basis beforehand to replace them. After Bob's operations, Alice introduces no disturbances on the particles sent out from Bob corresponding to the CTRL operations, but intercepts the particles sent out from Bob corresponding to the SIFT operations, uses the $Z$ basis to measure them and sends the resulted states to TP. In this way, Alice can easily deduce $B$ from the initial prepared states of the fake particles and her measurement results on the corresponding particles sent out from Bob before being detected. However, although Alice knows the positions of $Y'$ in $B$, she still cannot obtain $Y$ before being detected, because she has no access to $R_{B1}$ at the moment.

Apparently, the above intercept-resend attack from Alice introduces no error on the CTRL particles. In the following, we validate that when launching the above intercept-resend attack, Alice leaves her trace on the SIFT particles so that her attack behavior can be discovered by Bob and TP. For example, assume that an initial pair of particles corresponding to Alice and Bob's SIFT operations is prepared by TP in the state of $|\varphi^+\rangle$ and that one fake particle prepared by Alice is in the state of $|0\rangle$. Alice performs the above intercept-resend attack on the particle from TP to Bob and back to TP. TP executes Bell basis measurement on this pair of returned particles. When both Alice and Bob choose the SIFT operations, TP's measurement result is randomly in the state of $|\varphi^+\rangle$, $|\varphi^-\rangle$, $|\psi^+\rangle$ or $|\psi^-\rangle$; as a result, Alice's attack behavior is detected with the probability of $\frac{3}{4}$ when this pair of particles is chosen for security check.

Secondly, we analyze the measure-resend attack from Alice. Alice introduces no disturbances on the particles of $S_B$ sent out from TP corresponding to the CTRL operations, but intercepts the particles of $S_B$ sent out from TP corresponding to the SIFT operations, uses the $Z$ basis to measure them and sends the resulted states to Bob. After Bob's operations, Alice introduces no disturbances on the particles sent out from Bob corresponding to the CTRL operations, but intercepts the particles sent out from Bob corresponding to the SIFT operations, uses the $Z$ basis to measure them and sends the resulted states to TP. In this way, Alice can easily deduce $B$ from her measurement results on the particles of $S_B$ sent out from TP corresponding to the SIFT operations and the corresponding particles sent out from Bob before being detected. Unfortunately, although Alice is aware of the positions of $Y'$ in $B$, she still has no way to get $Y$ before being detected, due to lack of $R_{B1}$ at the moment.

Obviously, no error is induced by the above measure-resend attack from Alice on the CTRL particles. Then, we validate that Alice's above measure-resend attack disturbs the SIFT particles so that Bob and TP can detect her attack behavior. For instance, suppose that TP prepares an initial pair of particles corresponding to Alice and Bob's SIFT operations in the state of $|\varphi^+\rangle$. Alice imposes the above measure-resend attack on the particle from TP to Bob and back to TP. Without loss of generality, assume that after Alice measures the particle from TP to Bob with the $Z$ basis, this pair of particles is collapsed into $|0\rangle|0\rangle$. TP performs Bell basis measurement on this pair of returned particles. When both Alice and Bob choose the SIFT operations, Alice's

attack behavior is discovered with the probability of $\frac{1}{2}$ if this pair of particles is chosen for security check.

It is easy to know after similar analysis that, through his attack behaviors, the dishonest user Bob cannot know $X$ either before being detected, due to lack of $R_{A1}$ at the moment, although he knows the positions of $X'$ in $A$; and moreover, his attack behavior can be detected inevitably by Alice and TP.

(2) The participant attack from TP

In the proposed protocol, TP is supposed to be semi-honest, that is, she may try all possible means to get two users' private inputs but cannot be allowed to conspire with anyone. TP may adopt the following attack strategy: TP prepares $8n$ single particles in the $Z$ basis instead of $4n$ Bell states in Step 2, picks out half of these single particles to form $S_A$, and makes the remaining half particles compose $S_B$; afterward, TP sends the particles of $S_A$ ($S_B$) to Alice (Bob) one by one. After Alice's (Bob's) operations, when Alice and Bob tell TP the positions where they chose the CTRL operations, TP uses the $Z$ basis to measure the received particles whose positions Alice and Bob chose the SIFT operations. As a result, TP can decode out $A$ ($B$) from the initial prepared states of the SIFT particles in $S_A$ ($S_B$) and her measurement results on the corresponding particles in $S_A'$ ($S_B'$). In order not to be detected by Alice and Bob when checking the transmission security of CTRL particles, TP always announces Alice and Bob that her Bell basis measurement results on the CTRL particles of the same positions in $S_A'$ and $S_B'$ are identical to the initial Bell states prepared by herself; and in order not to be detected by Alice and Bob when checking the transmission security of SIFT particles, TP always publishes the initial prepared Bell states of the chosen positions and the corresponding Bell basis measurement results, which are correctly related to Alice's and Bob's corresponding unitary operations. In this way, TP's attack behavior can escape from being detected. In Step 6, TP hears the value of $r_{A1}^i \oplus k^i$ ($r_{B1}^i \oplus k^i$) from Alice (Bob). However, due to lack of $k^i$, TP still cannot decode out $r_{A1}^i$ ($r_{B1}^i$). As a result, TP cannot deduce $x_i$ ($y_i$) from $x_i'$ ($y_i'$) which is known by her ahead.

In addition, although TP knows $m_i = x_i \oplus y_i$, she still has no way to get the accurate values of $x_i$ and $y_i$.

## 5.Performance analysis

In this part, we compare the proposed protocol with previous SQPC protocols based on Bell states in Refs.[19-24]. The specific comparison results are shown in Table 2. Referring to Ref.[50], we define the qubit efficiency as $\eta = \frac{\lambda_s}{\lambda_q+\lambda_c}$, where $\lambda_s$, $\lambda_q$ and $\lambda_c$ represent the number of compared private bits, the number of consumed qubits and the number of classical bits for the classical communication, respectively. Here, we take no account of the classical resources consumed in the eavesdropping detection processes.

In the proposed protocol, both the length of $X$ and the length of $Y$ are $n$ bits, so it has $\lambda_s = n$. TP needs to prepare $4n$ initial Bell states; moreover, Alice and Bob share $K$ and $K_{AB}$ via the SQKD protocol in Ref.[16] beforehand, each of which requires Alice to generate $N = \lceil 4n(1+\delta) \rceil$ qubits randomly in $\{|0\rangle, |1\rangle, |+\rangle, |-\rangle\}$ and Bob to prepare $M$ qubits randomly in the $Z$ basis, where $|\pm\rangle = \frac{1}{\sqrt{2}}(|0\rangle \pm |1\rangle)$ and $M \geq N$. Hence, it has $\lambda_q = 4n \times 2 + 2N + 2M = 8n + 2N + 2M$. In addition, Alice (Bob) need tell TP the value of $r_{A1}^i \oplus k^i$ ($r_{B1}^i \oplus k^i$), where $i = 1,2,\dots,n$. As a result, it has $\lambda_c = n \times 2 = 2n$. Consequently, the qubit efficiency of the proposed protocol is $\eta = \frac{n}{8n+2N+2M+2n} \leq \frac{1}{26}$. Using the same method, we calculate the qubit efficiency of each of the protocols in Refs.[19-24].

According to Table 2, compared with the SQPC protocols of Refs.[19-24], the proposed SQPC protocol has the following merits: (1) it can be used to compare the equality of one quantum user's private input and one classical user's private input, but each of Refs.[19-24] is suitable for two classical users to compare the equality of their private inputs; (2) it only requires TP to carry out Bell basis measurements, but each of the protocols of Refs.[19,20,22,23] requires TP to perform both Bell basis measurements and $Z$ basis measurements; (3) it releases the classical user from quantum measurements, but each of the protocols of Refs.[19-24] requires the classical users to perform quantum measurements; (4) it doesn't need quantum entanglement swapping, but the protocol of Ref.[19] requires quantum entanglement swapping; (5) its qubit efficiency can be larger than that of each of the protocols in Refs.[19-24].

Table 2 Comparison results of our SQPC protocol and previous SQPC protocols based on Bell states

| | The protocol of Ref.[19] | The protocol of Ref.[20] | The second protocol of Ref.[21] | The protocol of Ref.[22] | The protocol of Ref.[23] | The protocol of Ref.[24] | Our protocol |
|---|---|---|---|---|---|---|---|
| Function | Compare the equality of two classical users' private inputs with the help of a semi-honest TP | Compare the equality of two classical users' private inputs with the help of a semi-honest TP | Compare the equality of two classical users' private inputs with the help of a semi-honest TP | Compare the equality of two classical users' private inputs with the help of a semi-honest TP | Compare the equality of two classical users' private inputs with the help of a semi-dishonest TP | Compare the equality of two classical users' private inputs with the help of a almost -honest TP | Compare the equality of one quantum user's private input and one classical user's private input with the help of a semi-honest TP |
| Type of TP | Semi-honest | Semi-honest | Semi-honest | Semi-honest | Semi-honest | Almost-honest | Semi-honest |
| Feature | Measure -resend | Measure -resend | Measure-Randomization -resend | Discard -resend | Measure -discard -resend | Measure -resend | Unitary operation-resend |
| TP's measurement operation | Bell basis measurements and $z$ basis measurements | Bell basis measurements and $z$ basis measurements | Bell basis measurements | Bell basis measurements and $z$ basis measurements | Bell basis measurements and $z$ basis measurements | Bell basis measurements | Bell basis measurements |
| The classical user' quantum measurement | $z$ basis measurements | $z$ basis measurements | $z$ basis measurements | $z$ basis measurements | $z$ basis measurements | $z$ basis measurements | No |
| Usage of SQKD or SQKA | No | Yes | Yes | Yes | Yes | Yes | Yes |
| Usage of quantum entanglement swapping | Yes | No | No | No | No | No | No |
| Usage of unitary operations | No | No | No | No | No | No | Yes |
| Usage of delay lines | No | Yes | Yes | No | No | No | Yes |
| Qubit efficiency | $\frac{1}{82}$ | $\frac{1}{60}$ | $\frac{1}{32}$ | $\frac{1}{48}$ | $\frac{1}{36}$ | $\frac{1}{58}$ | $\leq\frac{1}{26}$ |

## 6.Generalization of the proposed SQPC protocol into its counterpart of the collective-dephasing noise quantum channel

In the section, we generalize the above protocol into its counterpart of the collective-dephasing noise quantum channel.

The collective-dephasing noise can keep $|0\rangle$ stationary and turn $|1\rangle$ into $e^{i\phi}|1\rangle$, where $\phi$ is the collective-dephasing noise parameter changing along with time. [51] $|0_{dp}\rangle = |01\rangle$ and $|1_{dp}\rangle = |10\rangle$ are two logical qubits invariant towards the collective-dephasing noise. [51] $|\pm_{dp}\rangle = \frac{1}{\sqrt{2}}(|0_{dp}\rangle \pm |1_{dp}\rangle) = \frac{1}{\sqrt{2}}(|01\rangle \pm |10\rangle)$ are also immune to this kind of noise.[52] $Z_{dp} = \{|0_{dp}\rangle, |1_{dp}\rangle\}$ and $X_{dp} = \{|+_{dp}\rangle, |-_{dp}\rangle\}$ are two sets of corresponding logical measuring bases. In addition, the four logical Bell states of Eq.(2), [53] are also invariant towards this kind of noise. These logical Bell states can be distinguished by imposed with two Bell basis measurements on the 1st and the 3rd physical qubits, and on the 2nd and the 4th physical qubits, respectively. [53]

$$|\Phi^+_{dp}\rangle_{1234} = \frac{1}{\sqrt{2}}(|0_{dp}\rangle|0_{dp}\rangle + |1_{dp}\rangle|1_{dp}\rangle)_{1234}$$

$$= \frac{1}{\sqrt{2}}(|01\rangle|01\rangle + |10\rangle|10\rangle)_{1234}$$

$$= \frac{1}{\sqrt{2}}(|00\rangle|11\rangle + |11\rangle|00\rangle)_{1324}$$

$$= \frac{1}{\sqrt{2}}(|\varphi^+\rangle|\varphi^+\rangle - |\varphi^-\rangle|\varphi^-\rangle)_{1324},$$

$$|\Phi^-_{dp}\rangle_{1234} = \frac{1}{\sqrt{2}}(|0_{dp}\rangle|0_{dp}\rangle - |1_{dp}\rangle|1_{dp}\rangle)_{1234}$$

$$= \frac{1}{\sqrt{2}}(|01\rangle|01\rangle - |10\rangle|10\rangle)_{1234}$$

$$= \frac{1}{\sqrt{2}}(|00\rangle|11\rangle - |11\rangle|00\rangle)_{1324}$$

$$= \frac{1}{\sqrt{2}}(|\varphi^-\rangle|\varphi^+\rangle - |\varphi^+\rangle|\varphi^-\rangle)_{1324},$$

$$|\Psi^+_{dp}\rangle_{1234} = \frac{1}{\sqrt{2}}(|0_{dp}\rangle|1_{dp}\rangle + |1_{dp}\rangle|0_{dp}\rangle)_{1234}$$

$$= \frac{1}{\sqrt{2}}(|01\rangle|10\rangle + |10\rangle|01\rangle)_{1234}$$

$$= \frac{1}{\sqrt{2}}(|01\rangle|10\rangle + |10\rangle|01\rangle)_{1324}$$

$$= \frac{1}{\sqrt{2}}(|\psi^+\rangle|\psi^+\rangle - |\psi^-\rangle|\psi^-\rangle)_{1324},$$

$$|\Psi^-_{dp}\rangle_{1234} = \frac{1}{\sqrt{2}}(|0_{dp}\rangle|1_{dp}\rangle - |1_{dp}\rangle|0_{dp}\rangle)_{1234}$$

$$= \frac{1}{\sqrt{2}}(|01\rangle|10\rangle - |10\rangle|01\rangle)_{1234}$$

$$= \frac{1}{\sqrt{2}}(|01\rangle|10\rangle - |10\rangle|01\rangle)_{1324}$$

$$= \frac{1}{\sqrt{2}}(|\psi^-\rangle|\psi^+\rangle - |\psi^+\rangle|\psi^-\rangle)_{1324}. \quad (2)$$

The counterpart of the above protocol in the case of the collective-dephasing noise quantum channel can be described as follows:

Step 1: Alice (Bob) applies a random number generator to produce two $n$ -bit random number sequences $R_{A1} = (r^1_{A1}, r^2_{A1}, \dots, r^n_{A1})$ ( $R_{B1} = (r^1_{B1}, r^2_{B1}, \dots, r^n_{B1})$ ) and $R_{A2} = (r^1_{A2}, r^2_{A2}, \dots, r^n_{A2})$ ($R_{B2} = (r^1_{B2}, r^2_{B2}, \dots, r^n_{B2})$), where $r^i_{A1}, r^i_{A2} \in \{0,1\}$ ($r^i_{B1}, r^i_{B2} \in \{0,1\}$) and $i = 1,2,\dots,n$. Then, Alice (Bob) encrypts $X$ ( $Y$ ) with $R_{A1}$ ( $R_{B1}$ ) to obtain $X' = \{x'_1, x'_2, \dots, x'_n\} = \{x_1 \oplus r^1_{A1}, x_2 \oplus r^2_{A1}, \dots, x_n \oplus r^n_{A1}\}$ ( $Y' = \{y'_1, y'_2, \dots, y'_n\} = \{y_1 \oplus r^1_{B1}, y_2 \oplus r^2_{B1}, \dots, y_n \oplus r^n_{B1}\}$). They agree on beforehand that $R_{A2}$ and $R_{B2}$ are located in the same positions of $X'$ and $Y'$ to compose new sequences $A = \{a_1, a_2, \dots, a_{2n}\}$ and $B = \{b_1, b_2, \dots, b_{2n}\}$ , respectively. Afterward, they pre-share two $n$ -bit private keys, $K = \{k^1, k^2, \dots, k^n\}$ and $K_{AB} = \{k^1_{AB}, k^2_{AB}, \dots, k^n_{AB}\}$ , via the collective-dephasing noise resistant version of the SQKD protocol in Ref.[16], where $k^i, k^i_{AB} \in \{0,1\}$ and $i = 1,2,\dots,n$. Note that the SQKD protocol in Ref.[16] can be turned into its collective-dephasing noise resistant version as long as the following requirements are satisfied: $|0\rangle, |1\rangle, |+\rangle$ and $|-\rangle$ are substituted by $|0_{dp}\rangle$, $|1_{dp}\rangle$, $|+_{dp}\rangle$ and $|-_{dp}\rangle$, respectively; the $Z$ basis and the $X$ basis are substituted by the $Z_{dp}$ basis and the $X_{dp}$ basis, respectively. According to $K_{AB}$, they produce another $4n$ -bit private key $K'_{AB} = \{k'^1_{AB}, k'^2_{AB}, \dots, k'^{4n}_{AB}\} = \{k^1_{AB}, k^2_{AB}, \dots, k^n_{AB}, k^1_{AB}, k^2_{AB}, \dots, k^n_{AB}, k^1_{AB}, k^2_{AB}, \dots, k^n_{AB}, k^1_{AB}, k^2_{AB}, \dots, k^n_{AB}\}$.

Step 2: TP produces $4n$ logical Bell states randomly in $\{|\Phi_{dp}^{+}\rangle, |\Phi_{dp}^{-}\rangle, |\Psi_{dp}^{+}\rangle, |\Psi_{dp}^{-}\rangle\}$. Then, she divides all first logical qubits and all second logical qubits into sequences $S_A$ and $S_B$, respectively. Finally, TP sends the logical qubits of $S_A$ ($S_B$) to Alice (Bob) one by one. Notes that after the first logical qubit is sent to Alice (Bob), TP sends the next one only after receiving the prior one.

Step 3: On receiving $s_A^j$ ($s_B^j$) from TP, when $k_{AB}^{\prime j} = 0$, Alice (Bob) performs the corresponding composite unitary operation on it and then send the resulted logical qubit to TP; and when $k_{AB}^{\prime j} = 1$, Alice (Bob) reflects it back to TP. Here, $s_A^j$ ($s_B^j$) is the $j$th logical qubit of $S_A$ ($S_B$), $j = 1,2,\ldots,4n$. The composite unitary operation is performed according to the following rule: when $a_l = 0$ ($b_l = 0$), Alice (Bob) imposes $I_1 \otimes I_2$ on the $l$th SIFT logical qubit; and when $a_l = 1$ ($b_l = 1$), Alice (Bob) imposes $\sigma_1 \otimes \sigma_2$ on the $l$th SIFT logical qubit. Here, $l = 1,2,\ldots,2n$, and the subscript in each original unitary operation denotes the physical qubit performed with it. For convenience, $S_A$ ($S_B$) after Alice's (Bob's) operations is denoted as $S_A^{'}$ ($S_B^{'}$).

Step 4: After receiving the logical qubits from Alice and Bob, TP uses logical Bell basis to measure the logical qubits of the same positions in $S_A^{'}$ and $S_B^{'}$, and writes down the corresponding measurement results. Alice and Bob inform TP of the positions where they chose the CTRL operations. TP produces a bit sequence $C = \{c_1, c_2, \ldots, c_{2n}\}$ according to her logical Bell basis measurement results on the SIFT logical qubits of the same positions in $S_A^{'}$ and $S_B^{'}$: when her measurement result is identical to the initial prepared state, TP makes $c_l = 0$; otherwise, she makes $c_l = 1$. Here, $l = 1,2,\ldots,2n$. The relationships among different parameters corresponding to SIFT logical qubits of the same positions in $S_A^{'}$ and $S_B^{'}$ are listed in Table 3.

In order to check the transmission security of CTRL logical qubits, for the positions where Alice and Bob chose the CTRL operations, TP checks whether her measurement results on the CTRL logical qubits of the same positions in $S_A^{'}$ and $S_B^{'}$ are identical to the initial states prepared by herself. If there are no negative results, the protocol will be continued; otherwise, the protocol will be stopped.

Table 3 Relationships among different parameters corresponding to SIFT logical qubits of the same positions in $S_A^{'}$ and $S_B^{'}$

| Initial prepared states | $a_l$ | $b_l$ | Alice's composite unitary operation | Bob's composite unitary operation | TP's measurement result | $c_l$ |
|---|---|---|---|---|---|---|
| $\lvert\Phi_{dp}^{+}\rangle$ | 0 | 0 | $I_1 \otimes I_2$ | $I_1 \otimes I_2$ | $\lvert\Phi_{dp}^{+}\rangle$ | 0 |
| $\lvert\Phi_{dp}^{+}\rangle$ | 0 | 1 | $I_1 \otimes I_2$ | $\sigma_1 \otimes \sigma_2$ | $\lvert\Psi_{dp}^{+}\rangle$ | 1 |
| $\lvert\Phi_{dp}^{+}\rangle$ | 1 | 0 | $\sigma_1 \otimes \sigma_2$ | $I_1 \otimes I_2$ | $\lvert\Psi_{dp}^{+}\rangle$ | 1 |
| $\lvert\Phi_{dp}^{+}\rangle$ | 1 | 1 | $\sigma_1 \otimes \sigma_2$ | $\sigma_1 \otimes \sigma_2$ | $\lvert\Phi_{dp}^{+}\rangle$ | 0 |
| $\lvert\Phi_{dp}^{-}\rangle$ | 0 | 0 | $I_1 \otimes I_2$ | $I_1 \otimes I_2$ | $\lvert\Phi_{dp}^{-}\rangle$ | 0 |
| $\lvert\Phi_{dp}^{-}\rangle$ | 0 | 1 | $I_1 \otimes I_2$ | $\sigma_1 \otimes \sigma_2$ | $\lvert\Psi_{dp}^{-}\rangle$ | 1 |
| $\lvert\Phi_{dp}^{-}\rangle$ | 1 | 0 | $\sigma_1 \otimes \sigma_2$ | $I_1 \otimes I_2$ | $\lvert\Psi_{dp}^{-}\rangle$ | 1 |
| $\lvert\Phi_{dp}^{-}\rangle$ | 1 | 1 | $\sigma_1 \otimes \sigma_2$ | $\sigma_1 \otimes \sigma_2$ | $\lvert\Phi_{dp}^{-}\rangle$ | 0 |
| $\lvert\Psi_{dp}^{+}\rangle$ | 0 | 0 | $I_1 \otimes I_2$ | $I_1 \otimes I_2$ | $\lvert\Psi_{dp}^{+}\rangle$ | 0 |
| $\lvert\Psi_{dp}^{+}\rangle$ | 0 | 1 | $I_1 \otimes I_2$ | $\sigma_1 \otimes \sigma_2$ | $\lvert\Phi_{dp}^{+}\rangle$ | 1 |
| $\lvert\Psi_{dp}^{+}\rangle$ | 1 | 0 | $\sigma_1 \otimes \sigma_2$ | $I_1 \otimes I_2$ | $\lvert\Phi_{dp}^{+}\rangle$ | 1 |
| $\lvert\Psi_{dp}^{+}\rangle$ | 1 | 1 | $\sigma_1 \otimes \sigma_2$ | $\sigma_1 \otimes \sigma_2$ | $\lvert\Psi_{dp}^{+}\rangle$ | 0 |
| $\lvert\Psi_{dp}^-\rangle$ | 0 | 0 | $I_1 \otimes I_2$ | $I_1 \otimes I_2$ | $\lvert\Psi_{dp}^-\rangle$ | 0 |
| $\lvert\Psi_{dp}^-\rangle$ | 0 | 1 | $I_1 \otimes I_2$ | $\sigma_1 \otimes \sigma_2$ | $\lvert\Phi_{dp}^-\rangle$ | 1 |
| $\lvert\Psi_{dp}^-\rangle$ | 1 | 0 | $\sigma_1 \otimes \sigma_2$ | $I_1 \otimes I_2$ | $\lvert\Phi_{dp}^-\rangle$ | 1 |
| $\lvert\Psi_{dp}^-\rangle$ | 1 | 1 | $\sigma_1 \otimes \sigma_2$ | $\sigma_1 \otimes \sigma_2$ | $\lvert\Psi_{dp}^-\rangle$ | 0 |

In order to check the transmission security of SIFT logical qubits, Alice and Bob publish the positions where they performed the composite unitary operations according to $R_{A2}$ and $R_{B2}$, respectively. Afterward, they ask TP to publish the initial prepared states of these chosen positions and the corresponding measurement results, while TP requires them to publish the values of $R_{A2}$ and $R_{B2}$, respectively. Then, Alice, Bob and TP check whether the initial prepared states of these chosen positions, TP's corresponding measurement results and Alice and Bob's corresponding composite unitary operations are correctly related or not. If there are no negative results, the protocol will be continued; otherwise, the protocol will be stopped.

Step 5: TP discards the bits in $C$ related to $R_{A2}$ and $R_{B2}$ to obtain the new bit sequence $C' = \{c_1', c_2', \ldots, c_n'\}$. Note that the bits in $C'$ are related to $X'$ and $Y'$.

Step 6: Alice (Bob) informs TP of the value of $r_{A1}^i \oplus k^i$ ($r_{B1}^i \oplus k^i$), where $i = 1,2,\ldots,n$. Afterward, TP calculates $m_i = (r_{A1}^i \oplus k^i) \oplus (r_{B1}^i \oplus k^i) \oplus c_i'$ for $i = 1,2,\ldots,n$. TP concludes that $X \neq Y$ as long as she finds out $m_i \neq 0$ for certain $i$; otherwise, TP concludes that $X = Y$. Finally, TP tells Alice and Bob the final comparison result of $X$ and $Y$.

## 7.Conclusions

In this paper, we propose a novel SQPC protocol based on Bell states, which is secure against the attacks from both internal participants and an external eavesdropper. Our protocol only requires TP to perform Bell basis measurements, releases the classical user from quantum measurements and a quantum memory, and doesn't need quantum entanglement swapping. Our protocol exceeds the SQPC protocols of Refs.[19-24] in the following aspects: (1) our protocol can compare the equality of the private inputs from one quantum user and one classical user, but Refs.[19-24] cannot do this; (2) our protocol takes advantage over Refs.[19,20,22,23] on TP's quantum measurements, as each of Refs.[19,20,22,23] requires TP to perform both Bell basis measurements and $Z$ basis measurements; (3) our protocol defeats Refs.[19-24] on classical user's quantum measurements, as each of Refs.[19-24] requires the classical users to perform quantum measurements; (4) our protocol exceeds Ref.[19] in quantum entanglement swapping, as Ref.[19] requires it; (5) compared with Refs.[19-24], our protocol's qubit efficiency is higher. In addition, our protocol can be generalized into its counterpart of the collective-dephasing noise quantum channel.

As we know, there are also numerous kinds of noisy quantum channel. However, we only consider the collective-dephasing noise quantum channel in this paper. How to design SQPC protocols suitable for other kinds of noisy quantum channel is worthy of studying in the future.